\def \SAIT #1 #2 {{\em Mem.\ Soc.\ Astron.\ It.\/} {\bf #1}, #2}
\def \MESS #1 #2 {{\em The Messenger\/} {\bf #1}, #2}
\def \ASTRNACH #1 #2 {{\em Astron. Nach.\/} {\bf #1}, #2}
\def \AAP #1 #2 {{\em Astron. Astrophys.\/} {\bf #1}, #2}
\def \AAL #1 #2 {{\em Astron. Astrophys. Lett.\/} {\bf #1}, L#2}
\def \AAR #1 #2 {{\em Astron. Astrophys. Rev.\/} {\bf #1}, #2}
\def \AAS #1 #2 {{\em Astron. Astrophys. Suppl. Ser.\/} {\bf #1}, #2}
\def \AJ #1 #2 {{\em Astron. J.\/} {\bf #1}, #2}
\def \ANNREV #1 #2 {{\em Ann. Rev. Astron. Astrophys.\/} {\bf #1}, #2}
\def \APJ #1 #2 {{\em Astrophys. J.\/} {\bf #1}, #2}
\def \APJL #1 #2 {{\em Astrophys. J. Lett.\/} {\bf #1}, L#2}
\def \APJS #1 #2 {{\em Astrophys. J. Suppl.\/} {\bf #1}, #2}
\def \APSS #1 #2 {{\em Astrophys. Space Sci.\/} {\bf #1}, #2}
\def \ASR #1 #2 {{\em Adv. Space Res.\/} {\bf #1}, #2}
\def \BAIC #1 #2 {{\em Bull. Astron. Inst. Czechosl.\/} {\bf #1}, #2}
\def \JSQRT #1 #2 {{\em J. Quant. Spectrosc. Radiat. Transfer\/} {\bf #1}, #2}
\def \MN #1 #2 {{\em Mon. Not. R. Astr. Soc.\/} {\bf #1}, #2}
\def \MEM #1 #2 {{\em Mem. R. Astr. Soc.\/} {\bf #1}, #2}
\def \PLR #1 #2 {{\em Phys. Lett. Rev.\/} {\bf #1}, #2}
\def \PASJ #1 #2 {{\em Publ. Astron. Soc. Japan\/} {\bf #1}, #2}
\def \PASP #1 #2 {{\em Publ. Astr. Soc. Pacific\/} {\bf #1}, #2}
\def \NAT #1 #2 {{\em Nature\/} {\bf #1}, #2}

\documentstyle{memsait}
\input epsf.sty
\begin{opening}
\title{TESTING THE AGN MODELS FOR THE X-RAY BACKGROUND} 
\author{R. Gilli$^1,^2$}
\institute{$^1$Universit\`a di Firenze, Dipartimento di Astronomia, 
Largo E. Fermi 5, I-50125, Firenze, Italy\\
$^2$Astrophysikalisches Institut Potsdam, An der Sternwarte, 16, D-14486, 
Potsdam, Germany}
\date{} % DO NOT INSERT ANY DATE HERE !!!
\end{opening}

\begin{document}

%\oddpagefooter{\sf Mem. S.A.It., Vol. ??, ??}{}{\thepage}
%\evenpagefooter{\thepage}{}{\sf Mem. S.A.It., Vol. ??, ??}
\oddpagefooter{}{}{} % LEAVE AS IT IS !
\evenpagefooter{}{}{} % LEAVE AS IT IS !
\ 
\bigskip

\begin{abstract}
We will briefly examine the following three issues related to the AGN 
synthesis models for X-ray background: 1) the possibility that absorbed 
AGNs evolve faster than unabsorbed ones; 
2) the existence of the still debated population of luminous absorbed AGNs, 
the so-called QSO2s; 3) the behavior of the AGN density at a redshift 
above $\sim 3$.
\end{abstract}

\section{Introduction}

The diffuse X-ray background (XRB) above $\sim 1$ keV is commonly 
explained as the integrated emission of unobscured plus obscured AGNs 
(Setti \& Woltjer 1989; Comastri et al. 1995). 
Strong support on AGN synthesis models has been 
provided by ROSAT and Chandra, which resolved most of the 0.5-2 keV 
and 2-10 keV background, respectively (Hasinger et al. 1998; Mushotzky et 
al. 2000). ROSAT sources were mainly identified as AGNs (Schmidt et al. 1998),
while the identifications of the Chandra sample are not complete.
Nevertheless, from the first results AGNs seem to be the dominant population.

Recent synthesis models (Gilli et al. 1999;
Pompilio et al. 1999), suggested that the number of absorbed AGNs, with 
respect to the unabsorbed ones, would increase with the redshift. 
This is in agreement with the results of Reeves \& Turner (2000), 
who found some evidence for the percentage of X-ray absorbed QSOs 
(mainly, but not only, radio-loud objects) to be higher at high redshift 
than in the local Universe.

Results from ASCA and BeppoSAX surveys (Akiyama et al. 2000; 
Fiore et al. 2000) show that many blue, broad lined optical QSOs at a 
redshift above z$\sim$1 have hard X-ray spectra, where the hardening is 
likely to be produced by absorption, in agreement with 
Reeves \& Turner (2000). According to the unified schemes, absorbed 
objects should not show broad lines in the optical. In fact,
luminous AGNs with only optical narrow lines, optical QSO2s, are not 
commonly observed. However, from an X-ray point of view, several QSO2s have 
already been discovered, although the informations on these objects as a 
population are still lacking. 

At present the behavior of the AGN space density at high redshift is not clear.
From optical and radio surveys it was found that the AGN  
density declines beyond z$\sim3$ (e.g. Schmidt et al. 1995), 
while from soft X-ray surveys (Miyaji et al. 2000) a constant density of 
AGNs beyond z$\sim3$ is not ruled out. 
We will show that, as soon as optical identifications of Chandra 
faint sources are complete, this issue will be settled.

\section{Tests}

{\it A different evolution for absorbed and unabsorbed AGNs?}\\
Starting from the synthesis model described in Gilli et al. (1999) we
have tried to verify if a model where the ratio R of absorbed to unabsorbed 
AGNs increases with the redshift provides a better description of the 
XRB spectrum and source 
counts with respect to a standard model where R does not evolve.
We first considered a model with R(z)=4 at every redshift and at every
luminosity (model A) and then a model where the ratio increases 
from R(0)=4 to R(z$_{cut}$=1.4)=7 at every luminosity (model B). 
We tuned the parameters in the X-ray luminosity function (XLF) of 
Miyaji et al. (2000) to obtain two models giving the 
same $\chi^2$ on the hard XRB spectral fit, and then we checked 
the differences in reproducing the soft and the
hard counts (Fig.~1). Model B provides a better description of the data,
with an improvement significant at $>99\%$ confidence level with respect 
to model A. However, this result depends on some systematic
uncertainties in the data and a full statistical test including fits 
to as many constraints as possible (e.g. redshift and absorption 
distributions in flux limited samples) should be performed before drawing 
any secure conclusions.\\ \\
{\it The effect of QSO2s.}\\
In order to test the effect of removing high luminosity absorbed 
objects, the QSO2s, from the model,
following Gilli et al. (1999) we introduced in model B an exponential 
cut-off in the XLF of absorbed AGNs (with an $e$-folding de-absorbed 
luminosity of $2\times10^{44}$erg s$^{-1}$ in the 0.5-2 keV range), 
and then re-fitted the XRB spectrum by
tuning the parameters of the XLF (model C).
As shown in Fig.~2, QSO2s are necessary to reproduce the ASCA
counts at fluxes of $\sim 10^{-13}$ cgs (erg cm$^{-2}$ s$^{-1}$).\\ \\
{\it The AGN density at high redshift.}\\
In the previous calculations we simply assumed that the AGN density increases
from z=0 to z$_{cut}$=1.4 and then remains constant up to z=4.5. 
Above z=4.5 no more AGNs were assumed to exist. Now, starting from model B, 
we check two different assumptions on the high redshift AGN density.  
Following Schmidt et al. (1995) we first assume an exponential decrease in 
the AGN density above z=3 (model D), and then no decline of the AGN 
density up to z=10 (model E).
Both models provide a good fit to the XRB spectrum, soft 
and hard counts, and redshift distributions. Then, with the data available 
at present it is not possible to discriminate between the two alternatives.
However, as shown in Fig.~2, the predictions for the redshift distribution 
of the Mushotzky et al. (2000) sample are significantly different: the 
percentage of objects above z=4 expected from model D and E are 5\% and 26\%,
respectively. Therefore, as soon as the optical identifications of this 
sample are complete, the behavior of the AGN density at high redshift will 
be determined.
 
\begin{figure}
\vspace{-2cm}
\hspace{0.5cm}
\mbox{\epsfxsize=5.5cm  \epsffile{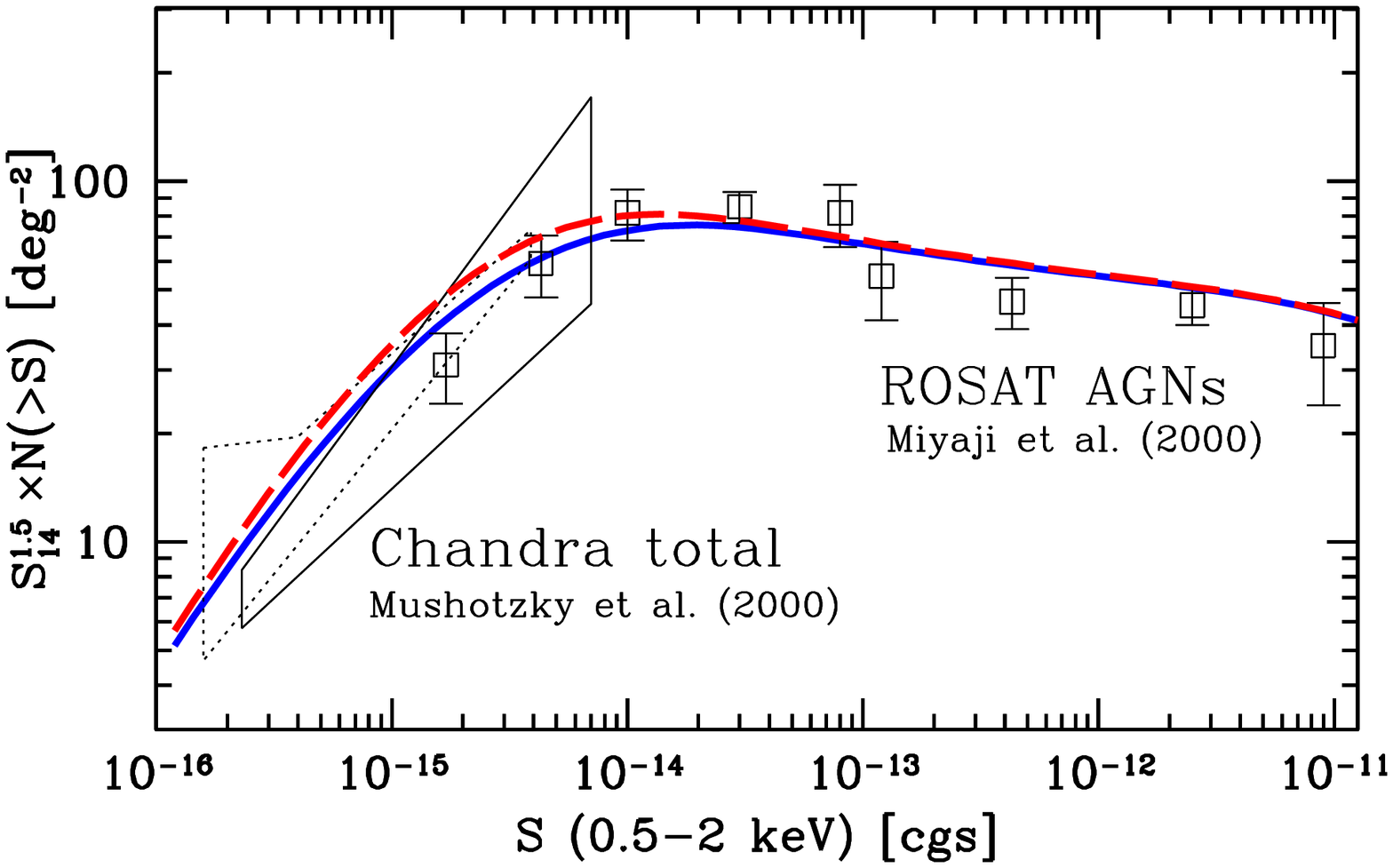}}
\hspace{1cm}
\mbox{\epsfxsize=5.5cm  \epsffile{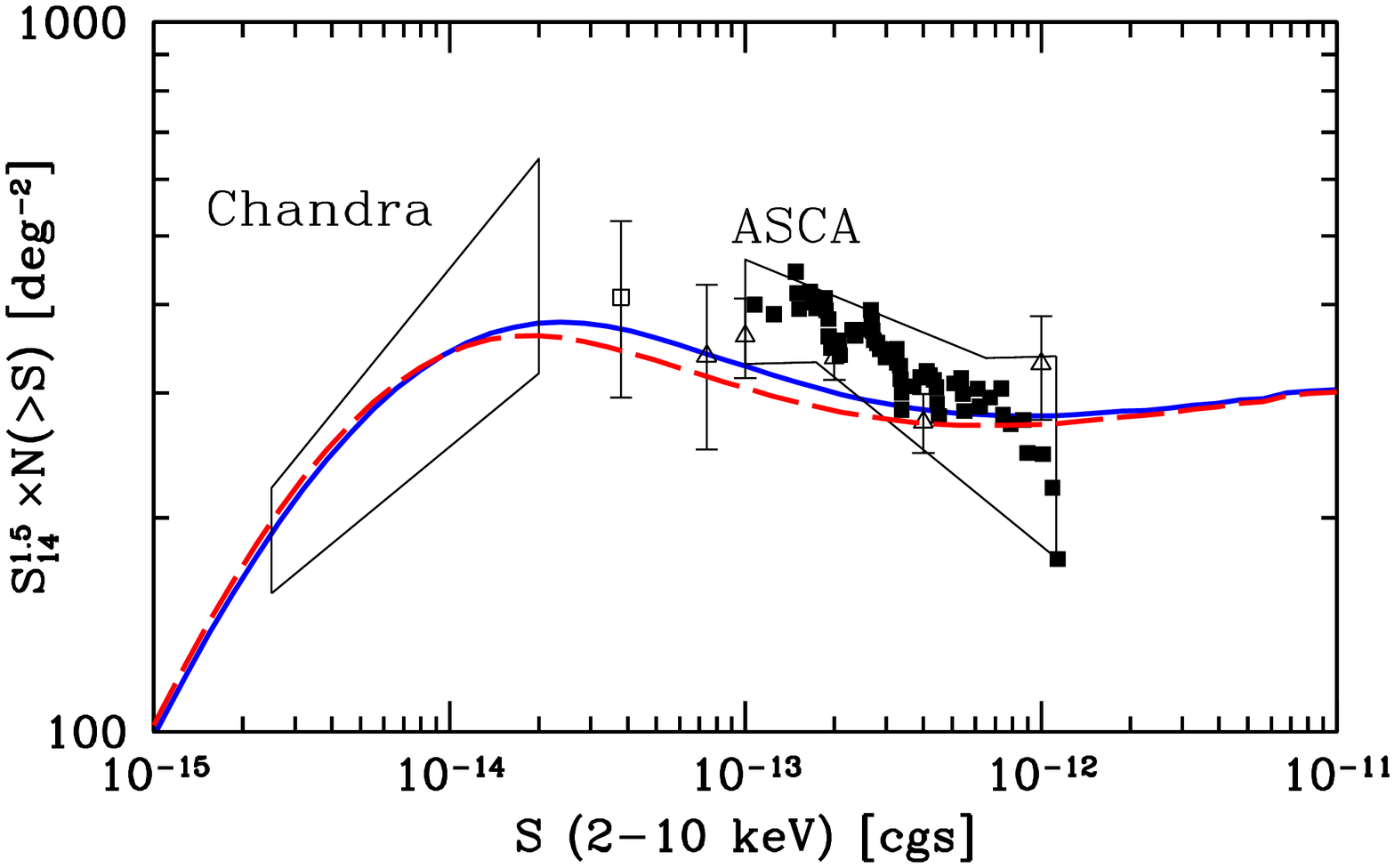}}
\vspace{-1cm}
\caption[h]{Comparison between the model predictions (dashed: model A; 
solid: model B) and source counts ({\it left}: soft counts;
{\it right}: hard counts). See text for details on the models.}
\end{figure}

\begin{figure}
\vspace{-2cm}   
\hspace{0.5cm}
\mbox{\epsfxsize=5.5cm  \epsffile{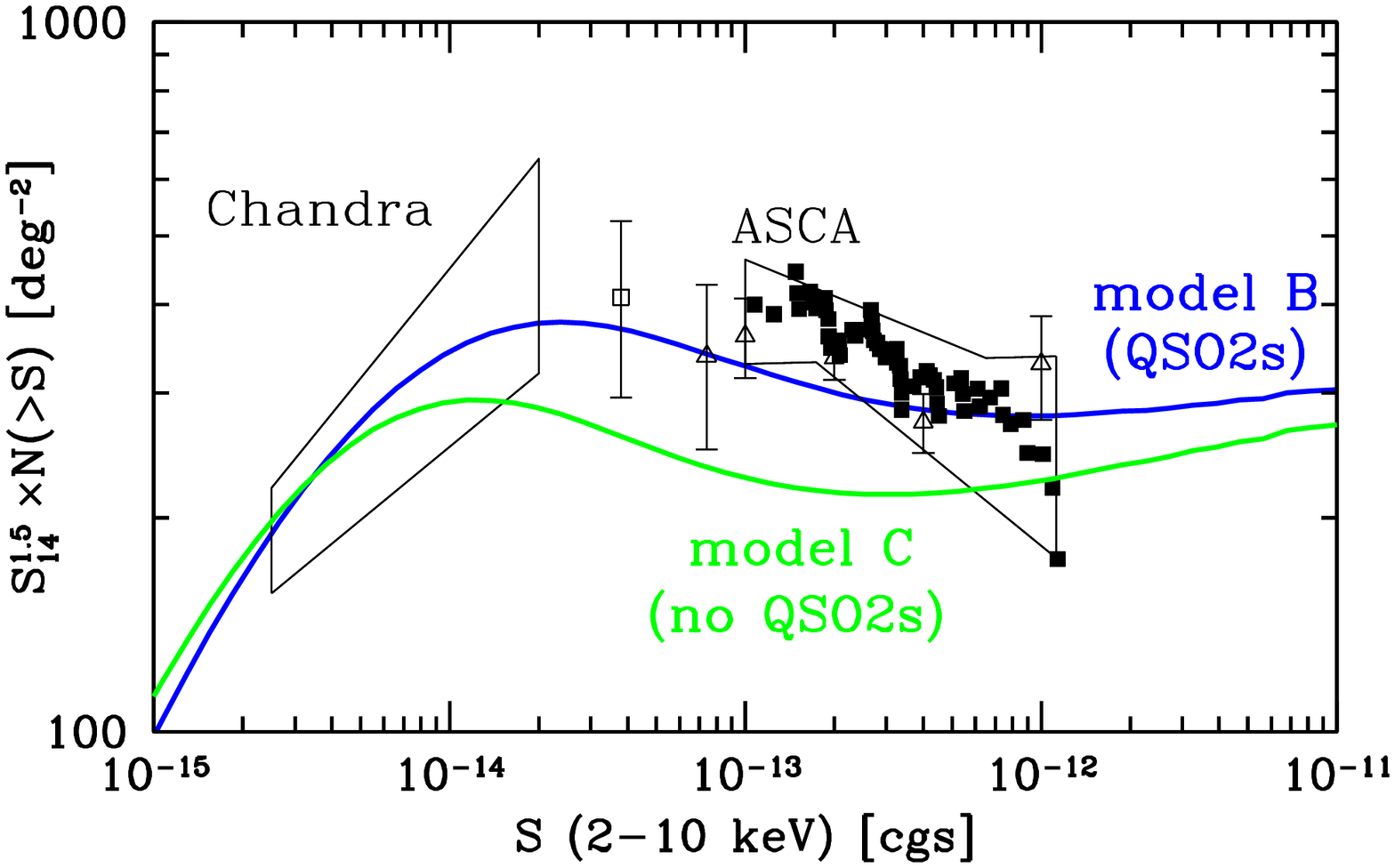}}
\hspace{1cm}
\mbox{\epsfxsize=5.5cm  \epsffile{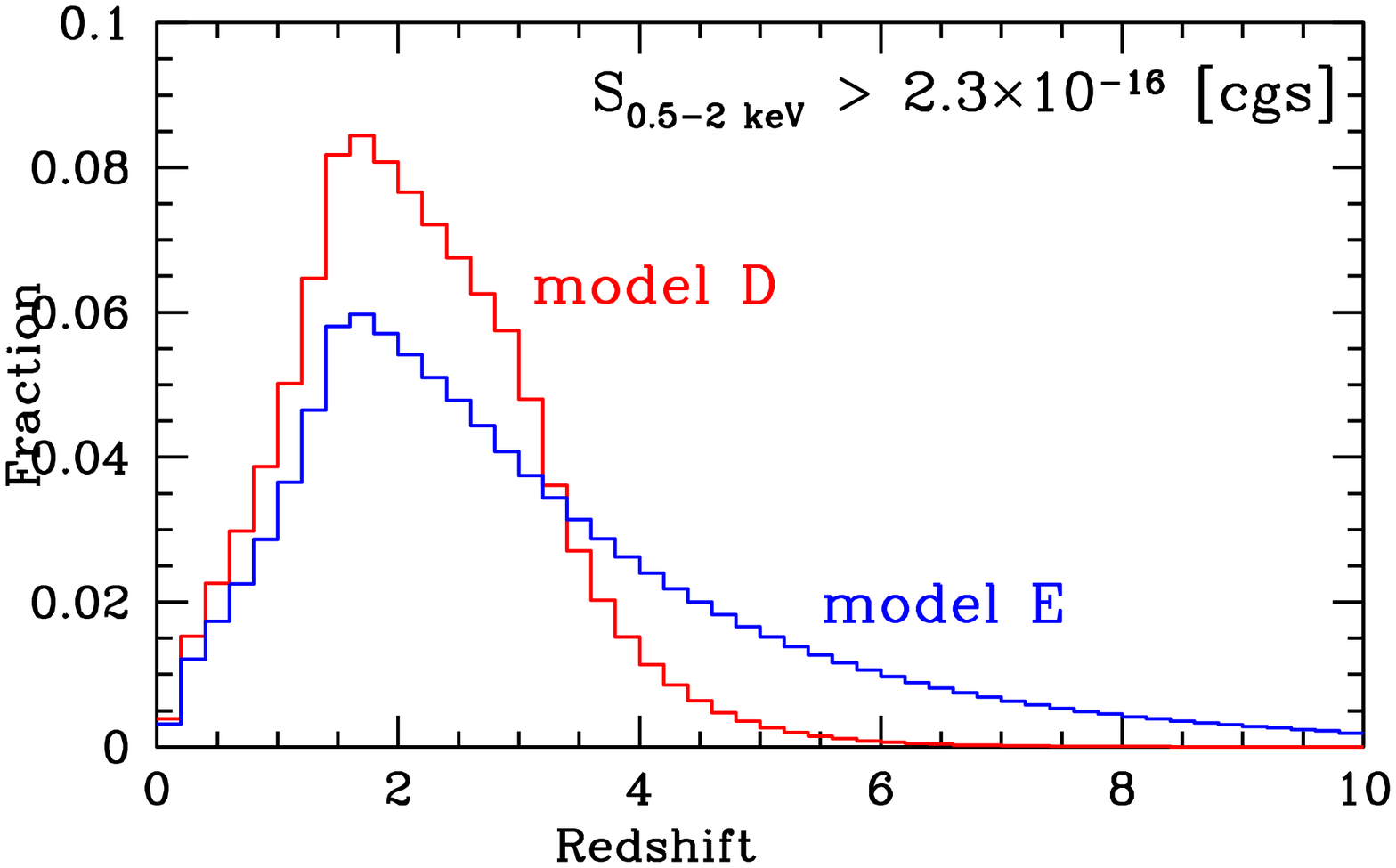}}
\vspace{-1cm}
\caption[h]{{\it Left}: Comparison between model predictions and hard counts. 
{\it Right}: Redshift distributions for the Mushotzky et al. (2000) 
sample as expected from model D and E.}
\end{figure}

\acknowledgements
This work has been done in collaboration with G. Hasinger and M. Salvati.
I thank A. Comastri for useful discussions.

% References. We avoided using the \bibitem commmand since we found it is
% somewhat platform-dependent. We also avoided using the \cite{keyword}
% command since we found it cumbersome. However, if you are an expert 
% LateX user you may use the various LateX tools for the references 
% provided they give the same printout formats of the examples given here.

\end{document}